# THIN CLIENT WEB-BASED CAMPUS INFORMATION SYSTEMS FOR FIJI NATIONAL UNIVERSITY

Bimal Aklesh Kumar

Department of Computer Science and Information Systems,
Fiji National University, Lautoka, Fiji
bimal.kumar@fnu.ac.fj

## ABSTRACT

*Fiji National University is encountering many difficulties with its current administrative systems. These difficulties include accessibility, scalability, performance, flexibility and integration. We propose a new campus information system, FNU-CIS to addresses these difficulties. FNU-CIS has the potential to provide wide range of the services for students and staffs at the university. In order to assist in the design and implementation of proposed FNU-CIS, we present an overview, software architecture and prototype implementation of our proposed system. We discuss the key properties of our system, compare it with other similar systems available and outline our future plans for research in FNU-CIS implementation.*

## 1. INTRODUCTION

Campus Information System (CIS) is a unified system that provides a single point of access to all secure administrative systems at higher education sector. These systems include, but are not limited to, student registration and enrolment, student and staff data, course work and exam information, program information, financial information, human resource information, accommodation, and many more as required by the institute [6]. CIS is a transaction processing system that serves at the operational level of the colleges and universities, it performs and records the routine transactions necessary to conduct its business. It also matches the structure, management tasks, instructional processes and special needs of the institution, like the traditional MIS, CIS integrates data from multiple sources to provide information stake holders need to make important management decisions [13].Traditionally CIS were mainly mainframe applications, since the late 1990's it has been changing and are fast adopted through the presence of a web medium as channel for accessing CIS with out any hassle upon viewing relevant information [10].

Fiji National University (FNU) was established in 2010 with the merger of six government owned tertiary institutions. It is a national institution, supporting the national effort for a stable economy and a literate population that is able to establish itself in the global community, while understanding and responding to the aspirations of individuals [7]. FNU has a network of thirteen campuses throughout the country. The objective of the FNU is to promote research and academic excellence for the welfare and needs of the communities in Fiji as well as communities in the region and abroad who wish to receive tertiary education of high quality at affordable cost.

Prior to the merger and formation of the Fiji National University (FNU) and due to the autonomous operations of these colleges, at least three different Campus Information Systems existed. The university is encountering considerable amount of difficulties with these systems.

These difficulties include:

- The system is only accessible to limited number of the university staff and not accessible to students.

- The performance of the system declines when it is required to handle high volume of transactions usually during the enrollment period.

- It requires expensive and complex upgrades as a condition of continued support and often theses changes do not bring the required improvement to the system.

- It is inflexible partly due to monolithic design and require IT staff to be involved with each change in business requirements.

- It involves large paper work to support key business activity such as enrollment, course adjustments, handling records of examinations, assessments mark, grades, and academic progression.

- The system does not interface well with other software applications used by the institute such as Finance, Human Resource Management and Timetabling systems.

In order to address the above difficulties, we propose thin client web-based FNU-CIS. This would be built using open source products and tools on modern code base with modern databases. FNU-CIS would have relatively clean separation between presentation, business logic, and data access layers, with solid data architectures and a well-defined set of business processes. It would be easily accessible to all the students and the staff of FNU through the local intranet or via World Wide Web. Data would be highly secured from unauthorized use. The system would serve user request with minimum response time. The design would be such where by subsequent modification will be limited as possible to least cost effect components and would not result in chain reaction of compensating modification, hence makes it easier to add more functionality in the future [2]. In this paper we describe a prototype implementation and software architecture of FNU-CIS using HTML, JSP, CORBA and MySQL. We furthermore discuss the key properties of our system and compare it with other systems available and finally outline our plans for future research.

## 2. RELATED WORK

The rapid growth of internet, intranets, extranets and other interconnected global networks in the 1990's dramatically changed the capabilities of information systems in organizations. Internet based and web-enabled systems are becoming common place in the operations of today's organizations [6]. Today's information systems are still doing the same thing they began doing 50 years ago, however what has changed is that we enjoy much higher level of integration of systems functions across applications and greater connectivity across both similar and dissimilar systems components.

The internet and other related technologies have changed the way businesses operate and people work, and how information systems support businesses processes, decision making and competitive advantage [7]. Today many organizations are using internet technologies to web enable business processes and to create innovative business applications.

The higher education administrational processes are undergoing a significant transition. The first major shift in these technologies occurred when they evolved from mainframe programs to client server solutions. Today, these technologies are more central then ever to colleges and universities, the changing technology landscape together with data reporting demands has compelled colleges and universities to evaluate their major software systems [10]. Now universities are mostly using centralized web based systems that is based on industry standard technology. These latest generation systems allow sharing of data easily with other systems and provide option to communicate important information to relevant stake holders [10].

Institutions are under constant pressure to demonstrate both willingness and capacity to incorporate the latest developments in CIS [3]. With younger generation having grown up with technology, students now expect to get information and do business with colleges and universities on the web [10]. Ease and convenience for the customer is essential for any university and we believe the student services on the web play a major part in this.

The institutions often have choice to carry in house development or buy an off the shelf packaged software [11]. Many colleges and universities have their own IT department that develops and manages the system. There are also many packaged software available such as People soft campus management systems provided by Oracle Corporation, Campus Management System provided by SAP etc.

Buying off the shelf challenges the notion that we are unique and have requirements that other similar organizations do not have [11]. Some of the benefits are software vendors attempt to incorporate the best practices from all of their customers into their products. New customers are able to take advantage of the experiences of others and the period of time between decisions to purchase through to the implementation would be shorter if the organization chose to develop the system in house. The major limitation for off the shelf software package is upgrade costs would be substantial, in addition to providing bug fixes new functionality upgrades are often required.

Custom developed software with proper analysis, design and implementation, the final product should meet the requirements defined by the user. User acceptance should be higher because of the input end users have in the design. Custom development often allows the organization to avoid 'big bang' implementation of a system through phased implementation. Some of the limitations could be that design development phases result in longer and larger projects.

The key architecture choice is between client server system and web-based system. The client-server products require a server for each site and use special software that must installed and maintained on each computer running the CIS [11]. In contrast web-based systems usually have centralized server that allows users to access data via a secured internet connection with out special software installations. The World Wide Web has emerged as powerful and appealing technology to utilize in the migration of main frame systems. Ease of development cost, platform independence and accessibility are some of the reasons that using the web is appealing. Continuing exponential growth in the availability of computing processor cycles, memory, storage and network bandwidth together with the rapid growth of World Wide Web have made it possible to develop modular and flexible systems.

## 3. OVERVIEW OF FNU-CIS

FUN-CIS has user interface designed using HTML and JSP, which simply collects data from the users and posts it to the server for processing, and then displays the processed data back to the user. The business logic would be provided by CORBA implemented application servers, and data would be permanently stored using My SQL database. The newly designed system will primarily have three main groups of users; students, academic staff and administration staff. Each of these users will be able to access the system using any standard web browser.

### 3.1. Student Users

- The current student and prospective students can apply to study at FNU using an online application form. Students are required to provide their personal details, proposed program of study, citizenship, funding details, qualifications and work experience, and attach the electronic copies of their results. The completed forms are then forwarded to respective department HOD's for their decision to approve or reject the application. If the application is approved students are notified with an offer letter which contains user name and password to use FUN-CIS, else the students are sent a letter stating their application has been declined stating the reasons.

- The students log into the system by using valid student id and password, upon verification of id and password they are directed to student menu page. Student menu page has several options students can to go to their profile, program details, graduation, enrollment, timetable, transcript, course work, class shares, finance and log out.

- The profile option provides them with choice to view and update their profile. Students are allowed to update their postal address, residential address, home phone number and mobile number.

- The program details option lists the units that students are required to complete for their program of study. The units are classified into three categories core units, major units and service units, the list also units to be completed.

- The graduation menu allows the students to fill the form for graduation. They will only be allowed to do this if they have completed all units required for their current program.

- The enrollment option allows the student to register for courses. This option also allows students to enroll for current and preceding terms of study. The students are required to select the term of study and the campus in which they want to study. All the units that are available in the particular campus for the selected term of study will be displayed. The system will list only those units that are required for students program and those units that the student has not yet completed. Once the student confirms the enrollment, the system checks if the students have passed the prerequisite the enrollments are approved if not they are sent to the HOD's for special approval. The students are given specified period of time in which they can change the units in which they are enrolled. Using this option student can also apply online to change their majors or programs of study.

- The timetable option allows the student view the class and final exam time table.

- The transcripts option lists all the units the student has completed so far. It displays the unit code, unit name, grade, campus, term of study and year.

- The course work option displays the student's course work for the current semester. The class shares option allows the student to access the class materials, FUN-CIS interfaces with other legacy system to service this request.

- The finance option allows the students to view their invoices and make fees payments online through credit cards.

- Finally the log out menu allows the users to terminate their session.

### 3.2. Academic Staff Users

The academic staffs include tutors, lecturers, senior lecturers and professors. They can log into the system using staff id and password, upon verification of ID and password they are directed to academic staff menu page. The academic staff menu page has options such as staff profile, enrolment, course work, class list and HR.

- Staff profile option allows the staff members to view and update their profile.

- The enrollment option allows academic staff members to enroll students for any unit that is activated for particular term of study and student is meeting the prerequisite.

- The student details option allows the staff to enter student ID and access the student profile and academic history of the student.

- The course work option allows the staff to submit the coursework for the units they are teaching. To minimize large data entry, staff can submit coursework using standard excel file.

- The class list option allows the staff to access the list of the students that are enrolled for a particular unit. The staff will be required to select unit code, term of study, year of study and campus.

- The HR option allows the users to access the HR system used by the institute.

- The log out option allows the academic staff users to terminate their session.

### 3.3. Administration Staff Users

The administration staff which includes department HOD's, heads of schools, academic services staff, deans are directed to the administration menu pages. The administration menu page has choices such as profile, student details, unit activation, applications, graduations, enrollment, and reports.

- The profile menu allows the admin staff to view and update their profile.
- The student details option allows the admin staff to enter student id and access the student profile or academic history of the student.
- The applications option allows the to view new applications and approve or reject these applications.
- The graduations option lists the details of those students who have applied to graduate and allows the admin staff to approve or reject this request.

- The unit activation option allows the HOD's to activate units that would be offered at each campus for a particular term of study.
- The program update option allows the admin staff to approve or reject the students request to change their program or majors.
- The reports option allows users to download various types of statistical reports for decision making.

The use-case diagram given below captures the requirements of the system.

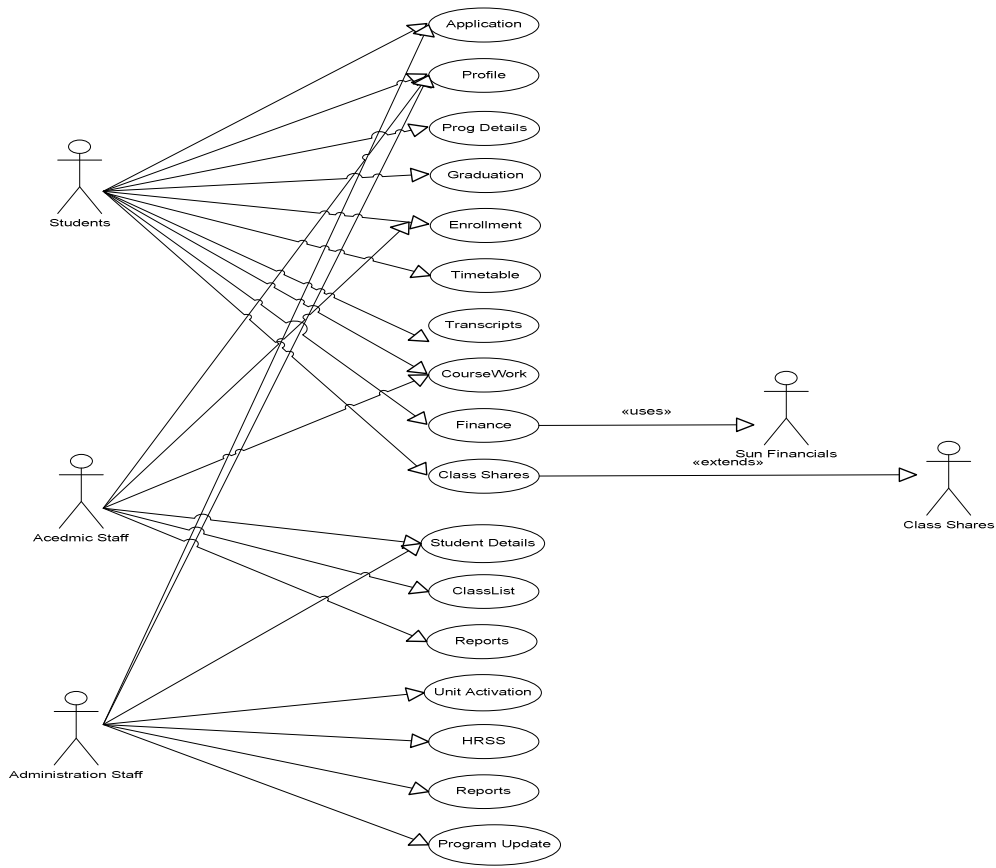

Figure 1. FNU-CIS Use Case Diagram

## 4. FNU-CIS SOFTWARE ARCHITECTURE

We have developed software architecture for implementing FNU-CIS for thin-client web-based applications. There are many platforms available to implement our system such as JINI, Microsoft .net and J2EE. We chose to implement our system using J2EE and CORBA. J2EE is based on java thus supports cross platform development and availability of world class open source free development environments like Eclipse and NetBeans lowers the overall development cost [8]. Similarly availability of open source application servers like CORBA along with database servers like MySQL allows both development and deployment to be extremely cost effective as compared to other proprietary application development platforms.

CORBA is extremely feature rich supporting numerous programming languages, operating systems and a diverse range of capabilities, such as transactions, messaging and security[14]. Many proprietary middleware technologies are designed with assumptions that developers will build applications using particular middle ware technology so they provide only limited support for integration with other technologies, in contrast CORBA was designed with the goal of making it easy to integrate with other technologies[12]. The flexible server side infrastructure of CORBA makes it feasible to develop servers that can scale from handling up to a number of objects to handling unlimited number of objects [14].

We designed multi-tier based software architecture for FNU-CIS. This includes client tier, web tier, application server tier and database tier. The client tier (web browser) is implemented using HTML, it displays data, collects input from the user and posts it back to the web server for processing which runs the JSP's to serve the request from the users. The web tier (web server), it includes JSP's and Java Beans which serve the request from browser client and generate dynamic content from them. Upon receiving the client request JSP's request from a Java Bean which in turn requests the information from CORBA implemented application servers. Once the Java Beans generate content, JSP's can query and display the content from Java Beans.

The application server tier is container for all the components. CORBA is used as the middleware which is implemented using Java language which has CORBA IDL mapping. Database tier is the backend of the system, My SQL is used to implement the database accessed via Java Database connectivity (JDBC).JDBC is an interface that allows java applications to connect to relational databases, when java applications interact with databases JDBC opens database connections and sends SQL commands to query the databases.

The newly designed system provides a single point of access for all the administrative applications used by the institute and would support activities that are not supported by the existing systems for all FNU campuses. The users run the web browser that accesses the FNU-CIS system. FNU-CIS web server provides single point of access to do business with the university online. The application server provides set of CORBA interfaces to communicate with web server and database server. The database holds all information such as student, staff, programs, units, timetable, grades, finance etc. Some colleges and departments may run some of their own software applications which may use quite different architectures and implementation technology. In the example below, CBHT may use a web server with Perl-implemented CGI scripts, C++ implemented application server and relational database. CAFF may use a J2EE-based architecture with J2EE server providing Java Server Pages (web user interface services) and Enterprise Java Beans (application server services), along with a relational database to hold data.

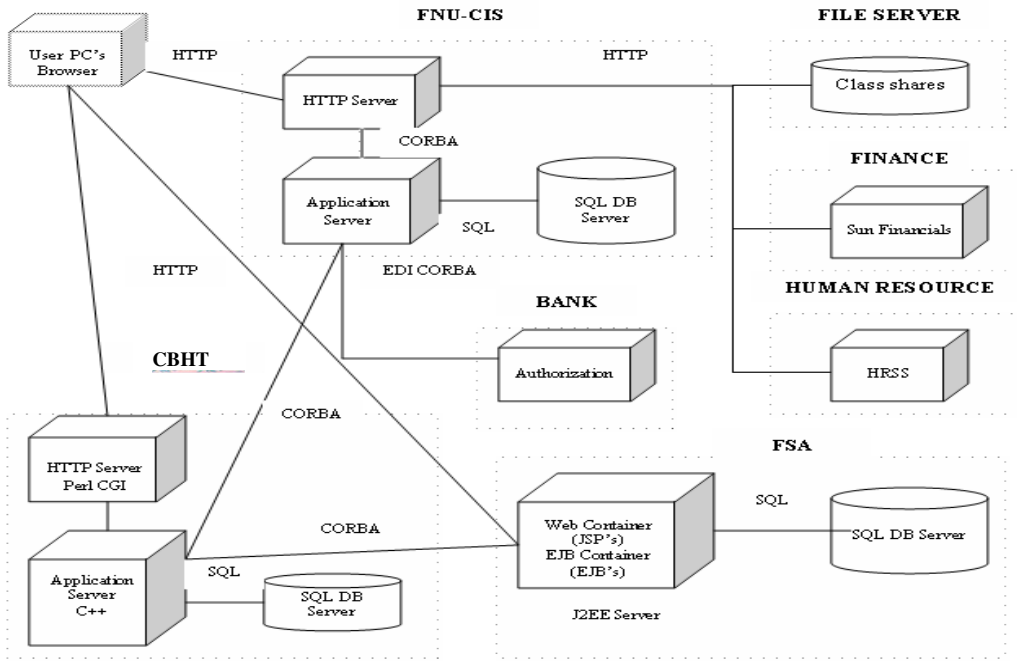

Figure 2. FNU-CIS Software Architecture

## 5. FNU-CIS PROTOTYPE EXAMPLE USAGE

In this section we present the prototype example usage of our FNU-CIS. We demonstrate one of the key features of our system, present the user interfaces and corresponding class interaction diagram based on multi-tier architecture design.

### 5.1. FNU-CIS LOGIN

Students login into the system using student id and password. The id and password are verified and if successful students are directed to the student menu page.

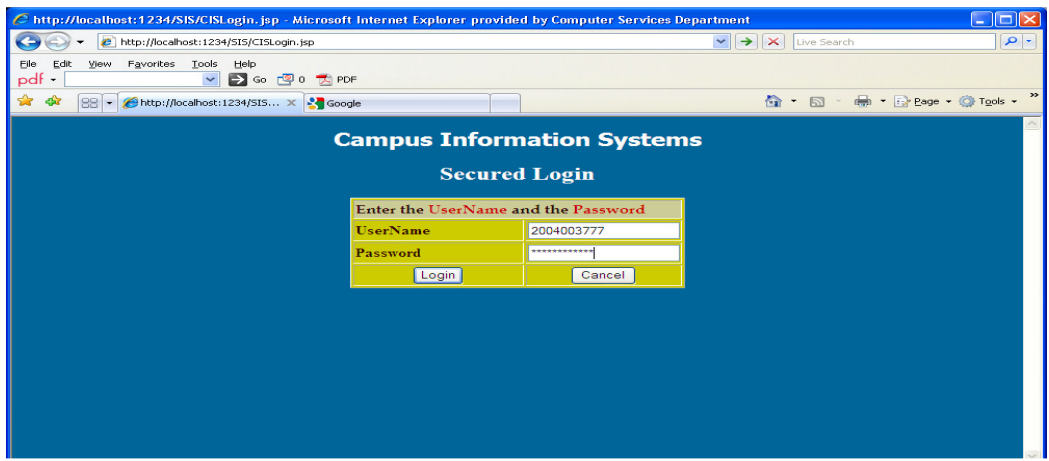

Figure 3. FNU-CIS Login Page

International Journal of Software Engineering & Applications (IJSEA), Vol.2, No.1, January 2011

Student menu page has several options students can to go to their profile, program details, graduation, enrollment, timetable, transcript, course work, class shares, finance and finally log out. Given below student selects profile option.

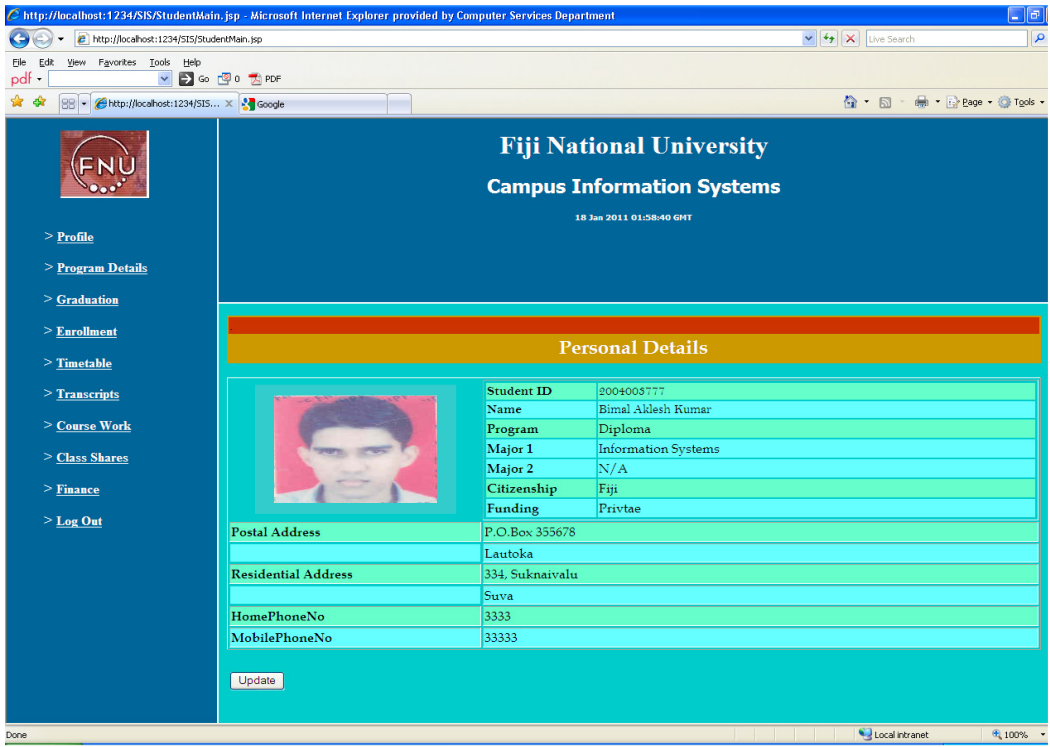

Figure 4. FNU-CIS Student Menu Page

The multi-tier class interaction diagram for FNU-CIS login is given below.

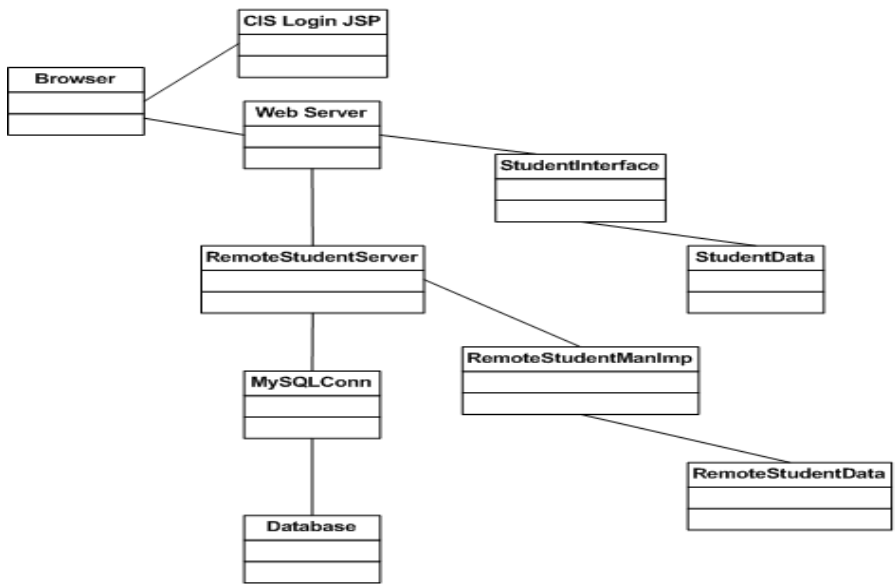

Figure 5. Class Interaction Diagram

The class details are as follows

CISLogin.jsp – is HTML and JSP implemented web page where users provide their username and password. The user name and password are assigned to a java bean (student data) and forwarded to student interface class.

Web server (Apache Tomcat) – provides the HTML and JSP implemented web pages. It serves the users http request for CISLogin.jsp page.

Student Interface - java implemented class that implements login functionality, it queries the data stored in student data object and invokes the remote login method implemented in RemoteStudentManager class.

StudentData – java implemented data class that stores student attributes and provides method to read and store student data.

RemoteStudentServer –java implemented CORBA application server that deploys remote classes such as RemoteStudentManagerImp.

RemoteStudentManagerImp –java implemented business class deployed in CORBA based RemoteStudentServer. It implements the login method which verifies user name and password from the database.

RemoteStudentData – java implemented data class that stores student attributes this class is used by RemoteStudentManagerImp to hold student data.

MySQLConn – java implemented class that provides functions to query data, update data, insert data and delete data from the database.

Database server – implemented using MySQL to provide physical data storage.

## 6. DISCUSSIONS

We compare FNU-CIS with other similar systems used by the universities in the region. We will use six comparison criteria to compare our system with other systems; accessibility, security, performance, flexibility, scalability and maintenance. The comparison criteria we used are based on key requirements for development of FNU-CIS.

- Accessibility - how often users can access the system.

- Security - protecting data from unauthorized use.

- Performance - how fast some aspect of a system performs under a particular workload.

- Flexibility - ability to accommodate change in business requirements with minimum modification.

- Scalability - the load of entity should not grow to an unmanageable size, the load should be distributed and shared.

- Maintenance - modification of a software product after delivery to correct faults, to improve performance or other attributes of the application.

International Journal of Software Engineering & Applications (IJSEA), Vol.2, No.1, January 2011

## 6. 1. CAMPUS INFORMATION SYSTEMS

There are many similar systems used by other institutes in region, we select three major systems and describe them, with which we will compare our proposed system.

### 6.1.1. ARTENA

Artena is a vender supplied system implemented in 2004 at Fiji Institute of Technology. It is a client-server based system requires installation for use by end users, thick clients communicate directly with database server. The system was only accessible for the administration staff. The system provided services such as enrollment, finance, student academic data management and statistical reports could also be generated from the system.

### 6.1.2. PREMIUM

Premium system was implemented at Fiji Institute of Technology in 2008, now is the core software application used by newly established FNU. It is a client server based system, thick clients communicate directly with database server. The system provides service such as enrollment, finance, maintains student academic data and provides statistical reports.

### 6.1.3. USP-CIS

University of the South Pacific is the oldest university in the south pacific region established in 1952. USP-CIS is a two tier web based system, staff and students can access the system using USP network or using World Wide Web. The system provides many services online that includes viewing student details, enrollment, graduation, finance, and academic data and also provides statistical reports.

The table given below provides a summary of comparison of our system with three other similar systems available

| System property | CAMPUS INFORMATION SYSTEMS | | | |
| --- | --- | --- | --- | --- |
| | *Artena* | *Premium* | *USP-CIS* | *FNU-CIS* |
| Accessibility | Only to limited staff during working hours | Only to limited staff during working hours | Available 24 x 7 to taff/students per authorization | Available 24 x 7 to staff/students per authorization |
| Maintenance | **High** maintenance as per changes in business requirements | **High** maintenance as per changes in business requirements | **Medium** maintenance as per changes in business requirements | **Low** maintenance staff can modify data values to suit business requirement changes |
| Security | Only authorised users and limited users have access to it | Only authorised users and limited users have access to it | Only authorised users and limited users have access to it | Only authorised users but available to all staff and students both. |
| Scalability | **Poor** due to its monolithic design | **Poor** due to its design | **Medium** due to its design | **High** due to its designed multi-tier architecture |
| Flexibilty | **Poor** due to its monolithic design | **Poor** due to its monolithic design | **Medium** – based on its design | **High** – based on its flexible design |
| Performance | **Poor** – often users not satisfied. | **Poor** – often users not satisfied | **Medium** – some times users are not satisfied | **High** – based on its flexible design. |

Table 1: Comparison of Campus Information Systems

23

## 6.2. CIS PROPERTIES COMPARISON

We provide a detail interpretation of each of the comparison criteria used.

### 6.2.1. ACCESSIBILITY

FNU-CIS would be deployed as thin client web-based application. Thin client applications display data collects input from the user and post it back to the server for processing. Thin clients don't require installation on user devices. It is generally easier to reach larger number of users in less heterogeneous locations through thin client architecture. It increases openness for use on multiple user devices with little memory and processing power. Thin client can arguably reach more devices such as (PDA, Mobile) due to its minimum requirements as many devices now have browsers [9].

### 6.2.2. SECURITY

All the systems given below have implemented security at various levels to secure data from unauthorized use. The proposed FNU-CIS would require user name and password that would be verified with user name and password stored on the system. FNU-CIS user interface is implemented using HTML and JSP that would validate data entered into the system. The software architecture is based on Object Oriented Design (OOD) where data and functions are wrapped as single unit called class. Data is not accessible to the outside world, only those functions that are wrapped in the class can access it thus makes the system more secured.

### 6.2.3. PERFORMANCE

It can also serve to validate and verify other quality attributes of the system, such as scalability and reliability [14]. FNU-CIS is deployed as thin client applications where as all other systems are deployed as thick client applications. Thick clients provide rich user interface that even allows the users to customize fonts and menus. FNU-CIS system would be faster as little data processing is done on the client, but instead data processing tasks are delegated to the supporting server. The client's primary responsibility is merely to display data, collect input from the user and post it back to the server.

### 6.2.4. FLEXIBILITY

The system is highly flexible, thin client's server plays a key part in implementing the business logic for the application and this can be centrally located and managed [8]. Changes can be easily rolled out to all users by changing the server side code, making deployments and updates simpler. However thick clients don't have this reliance on the server for basic navigation and processing logic. Software is easier to manage with thin client architecture because of centralization on server [9]. Developers can change code in a single place without needing to reinstall or update software on all user devices which may be geographically distributed.

### 6.2.5. SCALABILITY

Distributed systems easily expand and contract its resource pool to accommodate heavier or lighter loads [14]. A high performance application server is able to respond to increasing user needs. CORBA meets this requirement by internally coordinating the use of available memory and CPU resources through load-balancing and connection routing and pooling. The state failover capability makes it possible to run the same application on single CPU machines or high end SMP-clustered systems without changing the application.

### 6.2.6. MAINTENANCE

FUN-CIS would have lower maintenance as software application updates, virus scanning and patches can be executed on the server. Deployment costs are also reduced as thin clients can be remotely configured and do not need to be set up on individual machines. Break-fix simply requires replacing the thin client.

## 7. CONCLUSION

The proposed system would serve the students, academic and administration staff to carry out day to day business with the university online. This paper presented an overview, prototype example usage, software architecture and comparison of our proposed campus information systems for Fiji National University. The overview presented the functionalities together with the prototype example usage of the system. The software architecture described the designed multi-tier software architecture, separating presentation logic, web logic, business logic and data logic. Finally the system was compared with other available systems using the six key properties essential for development of an efficient Campus Information Systems. Future work would be carried on exploring further generalisation of our software architecture and evaluating the software system.

## ACKNOWLEDGEMENT

I would like to thank Dr. Sharlene Dai, senior lecturer at University of the South Pacific for supervising this research project and Fiji National University for providing financial support to carry out this research.

## AUTHORS

**BIMAL AKLESH KUMAR**

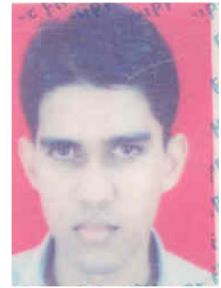

Bimal Aklesh Kumar is a lecturer at Fiji National University for past seven years in the department of Computing Science and Information Systems. His research interest includes software engineering, distributed systems and internet computing. Mr Kumar completed his BSC degree in Computing Science and Information Systems in 2002 from University of the South Pacific (USP). In 2003 he attained Microsoft Certified Professional (MCP) in Designing and Implementing Distributed Applications with Visual Basic 6.0. In 2010 he graduated with Postgraduate Diploma in Software Engineering from USP. At present Mr. Kumar is working on his Masters of Science thesis.